\documentclass[conference]{IEEEtran}
\IEEEoverridecommandlockouts

\usepackage{amsmath,amssymb,amsfonts}
\usepackage{graphicx}
\usepackage{textcomp}
\usepackage{xcolor}
\usepackage{bbm}
\usepackage{enumitem}
\usepackage{tabularx, booktabs}
\usepackage[style=ieee, maxcitenames=2, mincitenames=1, natbib=true, dashed=false, maxbibnames=99, minbibnames=99]{biblatex}
\usepackage{soul,color}
\usepackage{siunitx}
\usepackage{hyperref}
\usepackage[capitalise]{cleveref}
\usepackage{dsfont}
\usepackage{minted}
\usepackage[caption=false]{subfig}
\usepackage[linesnumbered, ruled, vlined]{algorithm2e}
\usepackage{float}
\usepackage{pifont}
\SetAlFnt{\small}

\crefname{algocf}{alg.}{algs.}
\Crefname{algocf}{Algorithm}{Algorithms}

\makeatletter
\renewcommand{\fnum@figure}{\textbf{Fig.~\thefigure}}
\makeatother

\usepackage{setspace}
\usepackage{multirow}

\usepackage[T1]{fontenc}
\usepackage{listings}
\DeclareCaptionStyle{ruled}{labelfont=normalfont,labelsep=colon,strut=off}
\floatstyle{ruled}
\newfloat{listing}{tb}{lst}{}
\floatname{listing}{Listing}

\crefname{lstlisting}{Prompt}{Prompts}

\lstdefinestyle{promptstyle_appendix}{
    backgroundcolor=\color{gray!8},
    basicstyle=\scriptsize\ttfamily\color{black!90},
    breakautoindent=false,
    breakindent=0pt,
    breakatwhitespace=true,
    breaklines=true,
    captionpos=b,
    keepspaces=true,
    showspaces=false,
    showstringspaces=false,
    showtabs=false,
    frame=single,
    rulecolor=\color{gray!40},
    framesep=3pt,
    frameround=tttt,
    framexleftmargin=6pt,
    xleftmargin=8pt,
    xrightmargin=8pt,
    tabsize=2,
    linewidth=1.\columnwidth,
    fontadjust=true,
    numbers=none,
    aboveskip=0.8\baselineskip,
    belowskip=0.8\baselineskip,
    columns=flexible,
    upquote=true,
    inputencoding=utf8,
    extendedchars=true,
    lineskip=-0.1pt,
    resetmargins=true,
    moredelim=[is][\color{blue!70!black}]{<slot>}{</slot>},
}

\usepackage{caption}

\begin{document}

\title{
    \LARGE \bf Robots in the City: Persona-Based Perceptions of Urban Robots
    \author{Keonhee Jang$^\star$, Kanghoon Lee$^\star$, Seokwoo Yun, Namwoo Kim, Yoonjin Yoon$^\dag$}

    \thanks{$^\star$Both authors contributed equally to this research.}
    \thanks{$^\dag$Corresponding author.}
    \thanks{K. Jang, N. Kim, and Y. Yoon are with the Spacetime Intelligence Laboratory, Korea Advanced Institute of Science and Technology (KAIST), Daejeon, South Korea. {\tt\small \{keonhee, namwoo, yoonjin\}@spacetime.kaist.ac.kr}.}
    
    \thanks{S. Yun is with Brighton Ai Group, Brisbane, QLD, Australia. {\tt\small ysu1516@gmail.com}. This work was done while he was with KAIST, Daejeon, South Korea.}
    \thanks{K. Lee is with the Systems Intelligence Laboratory (SILAB),
    KAIST,
    Daejeon, South Korea. {\tt\small leehoon@kaist.ac.kr} }

}
\maketitle

\begin{abstract}
Pedestrians increasingly encounter robots in urban public spaces without interacting with them, but how robots change the perceived safety of a street remains unclear.
We examine how robot presence and type affect perceived safety across urban contexts.
We generate 18 images of Seoul streets that vary in time of day, road class, and robot condition (no robot, a delivery robot, or a humanoid).
Two multimodal large language models (MLLMs), Qwen3.8-27B and Gemma4-26B, each conditioned on 500 synthetic personas, choose the safer image in each pair, and 25 human respondents complete the same task.
Both models prefer streets without robots and favor the delivery robot over the humanoid.
Robots are chosen more often at night and less often by older personas, whereas road class has smaller effects.
In the model rationales, the delivery robot appears as a physical obstacle, whereas the humanoid appears as either a threat or a source of protection.
Like humans, both models chose robots more often at night, but chose the humanoid over the delivery robot far less often, and Gemma showed a night effect for the humanoid about three times that of humans.
These results show that the effect of a robot on perceived safety depends on when it appears, and that persona-conditioned MLLMs capture only part of human judgments.
\end{abstract}

\section{Introduction}
Advances in embodied AI are bringing robots out of laboratories and into everyday urban public spaces.
People on the street often pass these robots without interacting with them directly, so robots must also be understood from the perspective of these pedestrians \cite{pelikan2024encountering,dobrosovestnova2025beyond}.
Even without interaction, a robot becomes part of the street scene \cite{pelikan2024encountering} and can change how safe the street feels, a long-standing topic in urban perception research \cite{naik2014streetscore,dubey2016deep}.
While prior studies has focused on the people who encounter robots without directly interacting with them \cite{dobrosovestnova2025beyond}, how robot presence and type shape the perceived safety of the surrounding streetscape remains unclear. Also, urban visual perception researches have rarely included robots.

Understanding this effect requires examining how people perceive a robot in a street environment: for example, an octopus on a farm would not fit its surroundings.
People recognize such an object less easily \cite{biederman1982scene} and look at it longer \cite{loftus1978cognitive}, which suggests that it takes more effort to process.
Objects that require more effort to process tend to be judged less favorably \cite{reber2004processing}, suggesting that an out-of-place object can make people feel uneasy.
A robot on a street is also an object in a scene, and whether it fits can depend on what it does and where and when it appears \cite{pelikan2025making}.
The same robot can thus fit one street or time of day but not another, changing how safe the street feels. To investigate this, we use street view images, a common stimulus in urban perception research \cite{naik2014streetscore,dubey2016deep}, and generate variants of each scene that differ in the robot and the time of day \cite{ye2025generated,dai2026urban}.
This allows us to examine how the urban context of a robot shapes its perceived safety.

We evaluate these images through pairwise comparisons of perceived safety, which allow controlled variants to be compared directly without relying on absolute ratings \cite{gu2026survey,salesses2013collaborative}.
We conduct these comparisons at scale using multimodal large language models (MLLMs) conditioned on persona descriptions, building on prior work on simulated responses \cite{aher2023using,argyle2023out} and visual reasoning in urban environments \cite{ro2025well}.
This allows us to examine whether the same visual changes are judged differently across personas.
We also collect the same judgments from human respondents to assess how closely the models align with human perception.
Because persona conditioning does not necessarily reproduce human heterogeneity \cite{hu2024quantifying}, we use persona variation as an analytical factor rather than a direct proxy for human populations.

In summary, our main contributions are as follows:
\begin{itemize}
    \item We examine how robot presence and type shape perceived safety across urban contexts.
    \item We compare human and persona-conditioned MLLM judgments, identifying where the models align with or depart from humans.
    \item We analyze the rationales behind model choices and how they shift with context.
\end{itemize}

\section{Related Work}

\textbf{Urban streetscape perception.}
Street view images have been widely used to study perceived safety \cite{naik2014streetscore}, broader subjective qualities of urban environments \cite{salesses2013collaborative,dubey2016deep}, and pedestrian-centered perceptions such as visual walkability \cite{li2022measuring,xiao2025urban}.
Recent work has examined how these judgments vary across observer groups \cite{cui2023analysing} and how visual AI can characterize perceptions of natural elements in urban scenes \cite{lefosse2025feeling}.
Prior studies have relied on comparisons across real-world street scenes, making it difficult to isolate the effect of a single visual feature.
More recent work has begun to address the issue by using generative transformations to create paired street view images with controlled changes, such as nighttime appearance \cite{ye2025generated}, and to compare perceived safety between matched daytime and nighttime conditions \cite{dai2026urban}.
Recent work has also evaluated vision-language models (VLMs) for spatial reasoning directly from street view images \cite{ro2025well}.
We extend this controlled visual comparison setting to examine how robot presence and type shape perceptions of urban pedestrian spaces.

\textbf{Robots in shared public space.}
Research on robots in public spaces shows that local context, physical environment, and social activity influence how people perceive robots \cite{pelikan2025making}.
These experiences are not limited to direct interaction, as people can perceive robots simply by seeing them in the surrounding environment \cite{dobrosovestnova2025beyond}.
Delivery robots are a common example because they operate on sidewalks and other pedestrian spaces alongside people \cite{abrams2021theoretical}.
When robots move through these spaces, pedestrian comfort and social constraints become important considerations for human-aware navigation \cite{kruse2013human}.
For sidewalk delivery robots, narrow paths and path intersections can make interactions more difficult \cite{gehrke2023observed}.
Environmental conditions can also affect how much distance people prefer to keep from robots \cite{nahum2026advancing}.
We instead examine how robot presence and type shape judgments of pedestrian space before motion or interaction occurs.

\textbf{Persona-conditioned models for subjective judgments.}
Large language models (LLMs) have been increasingly explored for simulating responses in human-subject studies \cite{aher2023using} and for generating persona-conditioned responses that reflect variation across individuals \cite{argyle2023out}.
In Human-Robot Interaction (HRI), LLMs have also been used as zero-shot human models, although their predictions can be sensitive to prompting and limited in reasoning \cite{zhang2023large}.
Recent work has also used generative agents to evaluate urban qualities such as safety and liveliness from street view images \cite{verma2026generative}.
However, persona conditioning does not necessarily reproduce human heterogeneity.
Its benefits can be limited when persona attributes are weakly related to the target judgments \cite{hu2024quantifying}, and matching empirical response distributions remains an open challenge \cite{meister2025benchmarking}.
We therefore interpret persona-conditioned MLLM outputs as simulated judgments rather than direct observations of human perception.

\section{Experimental Design}

\renewcommand{\lstlistingname}{Prompt}
\crefname{lstlisting}{Prompt}{Prompts}

\subsection{Scene and Condition Generation}

We collect three street scenes in Seoul from Mapillary \cite{neuhold2017mapillary}, one from each of three levels of the functional road hierarchy in Korean urban planning \cite{molit_road_classification}: primary arterial, secondary arterial, and local roads.
Each scene serves as a fixed setting for generating different robot and time of day conditions.

For each scene, we generate six visual conditions using GPT-5.6 Sol \cite{openai2026gpt56sol} by varying time of day and robot condition.
We first remove all people and vehicles, then set the time of day to a clear summer day or midnight, and reinsert people to reflect typical activity for each time and road class.
Each of these day and night images is then created in all three robot conditions, with no robot, a delivery robot, and a humanoid, yielding 18 images in total ($3$ road classes $\times$ $2$ times of day $\times$ $3$ robot conditions). 
All 18 images are provided in the \cref{app:stimuli}.
These variants preserve the original street layout, so that perceived safety can be compared across times of day, road classes, and robot types.

\begin{lstlisting}[caption={Prompt for persona-conditioned comparison
}, label={lst:annotation_prompt}, style=promptstyle_appendix, float=t, escapeinside={(*}{*)}]
(*\textbf{SYSTEM PROMPT}*)
(*\textbf{[Role]}*) You are the person described in the respondent card below.
(*\textbf{[Robots]}*) Two kinds of robots may appear. The delivery robot is a small, wheeled robot that helps people by delivering parcels to homes and bringing food orders from shops and restaurants. The humanoid is a two-legged, person-shaped robot that helps people by patrolling the street for security and assisting pedestrians who need help getting around. Whether a robot's presence is safe is for you to judge from the image.
(*\textbf{[Guide]}*) The two images show the same place at the same time. What differs between them is the robot: whether a robot is there, and if so, whether it is a delivery robot or a humanoid. Decide which image looks safer by thinking about how that difference affects your safety while walking here: what the robot is there to do, its size and shape, and where it is. A robot may make a place feel less safe or safer; judge that from the image. How much this matters depends on you, the person on the card. Draw on everything the card says about you: your gender, age, occupation and daily routine, household income, and where you live. Think about how your situation shapes walking in this place with or without that robot. Choose one image, and in your reason say what about the robot, or its absence, mattered most to you and why.
(*\textbf{[Format]}*) Output JSON only, with "reason" and "choice". Choose exactly one of the two images; ties are not allowed.

(*\textbf{USER MESSAGE}*)
(*\textbf{[Persona]}*) [Respondent card] <slot>{persona_card}</slot>
(*\textbf{[Context]}*) Imagine moving through the space in the image the way you usually do. Answer based on what is visible in the image and your own life experience.
(*\textbf{[Images]}*) <slot>{left_image}</slot> <slot>{right_image}</slot>
(*\textbf{[Task]}*) Which place looks safer? Write, in Korean, one or two sentences giving the main reason the image you chose looks safer. Answer choice with exactly "left" or "right". There is no tie.

(*\textbf{OUTPUT}*)
{"reason": "...", "choice": "left" | "right"}
\end{lstlisting}

\subsection{Persona-Conditioned MLLM Evaluation}

To examine whether evaluations vary across personas, we sample 500 synthetic personas from the Nemotron-Personas-Korea dataset \cite{nvidia2026nemotronkorea}, balancing the distributions of gender, age, household income, and residential region.
The resulting personas include 250 female and 250 male personas, 250 from the capital region and 250 from other regions, 166--167 in each of three income groups, and 83--84 in each of six age groups, from the 20s or younger to the 70s or older.

We conduct the evaluations using 4-bit quantized versions of two open-weight MLLMs, Qwen3.8-27B \cite{qwen2026qwen38_27b, redhat_qwen38_int4} and Gemma4-26B \cite{gemmateam2026gemma4, cyankiwi_gemma4_awq}.
Both models are served locally with vLLM \cite{kwon2023efficient} at a temperature of 0.7 for Qwen and 1.0 for Gemma, with thinking mode disabled.
In each trial, the model is shown a persona card and a pair of images depicting the same scene at the same time of day but with different robot conditions.
The left and right positions of the two images are randomized.
The persona card lists gender, age, occupation, household income level, and residence in Korean, followed by a one-line introduction and a short description of household income, as detailed in \cref{app:personas}.
The system prompt is the same for all inferences and includes a brief description of each robot, as shown in \Cref{lst:annotation_prompt}.
The model first writes a one- or two-sentence rationale in Korean and then chooses the image that looks safer, without ties.

Each persona evaluates the three robot pairs in each of the six combinations of road class and time of day, yielding 18 trials per persona and 9,000 original responses per model.
For consistency assessment, each persona repeats one trial per robot pair with both the original and swapped  image positions, yielding 3,000 additional responses per model.
Original responses are used for all analyses, and repeated responses are used only to assess consistency.

\subsection{Rationale Coding}
\label{sec:method_rationale}

To compare the reasoning expressed in model responses across conditions, we categorize the rationales from the 17,998 original responses (9,000 from Gemma and 8,998 from Qwen, after excluding two invalid Qwen responses).
Two authors review 100 randomly sampled rationales and identify five topics: (1) collision or path obstruction, (2) threat or unfamiliarity, (3) openness, visibility, or lighting, (4) surveillance, accompaniment, or protection, and (5) robot function or trust in technology.
Claude Opus 5.5 \cite{anthropic2026opus55} then codes each rationale up to three topics it mentions \cite{gilardi2023chatgpt}.
To assess reliability, a human annotator codes a random sample of 150 rationales, 25 for each of the six combinations of MLLM and robot pair, without seeing Claude's labels.
The agreement between the human annotator and Claude and further details of the coding procedure are provided in \cref{app:coding}.

\subsection{Human Evaluation}
\label{sec:method_human}

We conduct a parallel human survey to provide a human reference for the model evaluations.
We recruit 25 respondents (12 female and 13 male).
They participate voluntarily without compensation and give informed consent before completing the survey.
First, respondents report their gender and age group and read the same robot descriptions given to the models.
Next, they complete the same 18 trials as the models, selecting the image perceived as safer without ties, with randomized trial order and image position.
Finally, respondents report their education and household income and can describe what mattered most in their safety judgments.

\subsection{Analysis}
\label{sec:method_analysis}

We report the win rate as the share of trials in which the target condition is judged safer \cite{salesses2013collaborative}.
The target is the robot when it is compared with the empty street and the humanoid when the two robots are compared.
While we use all 500 personas for the main analyses, we match each human respondent with 10 personas of the same gender and age group for the human comparison.
We compute 95\% confidence intervals using a cluster bootstrap with 2,000 resamples of respondents or personas.
Differences by time of day and gender are expressed as Cohen's $h$, which places the gap between two win rates on a common scale, with 0.2, 0.5, and 0.8 marking small, medium, and large effects \cite{cohen1988statistical}.
Response consistency is the share of repeated trials in which a model chooses the same condition as before.

\section{Results}

\label{sec:results}

\subsection{Perceived Safety across Urban Contexts}
\label{sec:res_context}

\begin{figure}[t]
  \centering
  \includegraphics[width=\columnwidth]{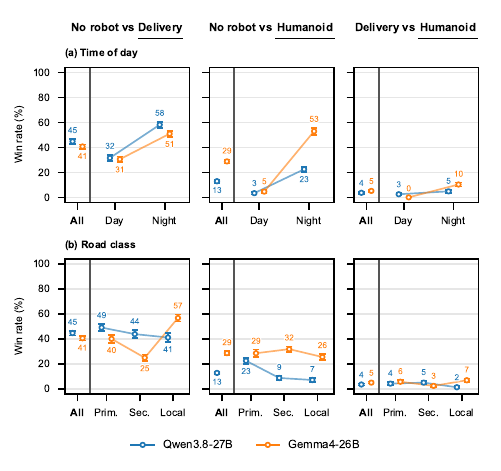}
  \caption{Win rate by urban contexts. Rows show the two conditions: (a) time of day and (b) road class, and columns show robot pairs.}
  \label{fig:context}
\end{figure}

\begin{figure}[t]
  \centering
  \includegraphics[width=\columnwidth]{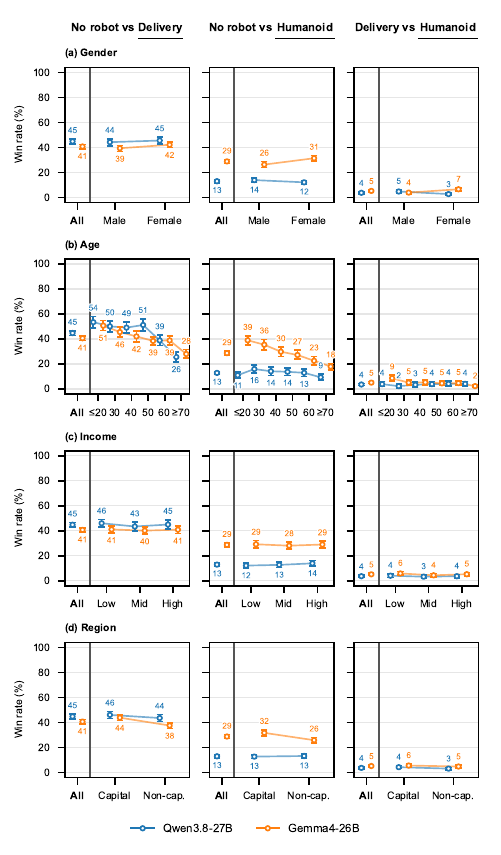}
    \caption{
    Win rate by persona attribute. Rows show the four attributes: (a) gender, (b) age, (c) income, and (d) region, and columns show robot pairs.
    }
  \label{fig:persona}
\end{figure}

\begin{figure*}[!t]
  \centering
  \includegraphics[width=\textwidth]{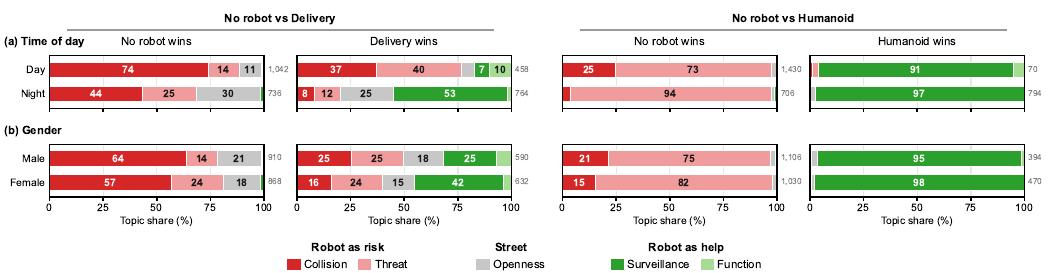}
  \caption{Share of rationale topics stated by Gemma4-26B for each choice, by (a) time of day and (b) persona gender, grouped by whether they describe the robot as a risk or a help, or the street. Shares within each bar sum to 100\%, and numbers to the right give the number of responses.}
  \label{fig:rationale}
\end{figure*}

We first examine how perceived safety changes with robot presence and type across urban contexts.
As shown in \Cref{fig:context}, both models chose the no-robot condition over the delivery robot and humanoid in most trials.
These results suggest that robot presence can reduce perceived street safety.
The delivery robot was chosen in 45\% and 41\% of trials by Qwen and Gemma, respectively, against the no-robot condition, whereas the humanoid was chosen in 13\% and 29\%.
Against the delivery robot, the humanoid was chosen in only 4\% and 5\% of trials. Overall, both models ranked the no-robot condition first, the delivery robot second, and the humanoid last.

\Cref{fig:context}a shows that the win rate of the delivery robot against the no-robot condition increased from 32\% to 58\% for Qwen and from 31\% to 51\% for Gemma.
The humanoid also gained against the no-robot condition, from 3\% to 23\% for Qwen and from 5\% to 53\% for Gemma.
Against the delivery robot, however, the humanoid remained rarely chosen at night (5\% and 10\%).
Overall, both models chose robots more often at night, but still favored the delivery robot over the humanoid.

We also examine how judgments vary across road classes, as shown in \Cref{fig:context}b.
For Qwen, win rates decreased from the primary to the local road, from 49\% to 41\% for the delivery robot and from 23\% to 7\% for the humanoid.
In contrast, Gemma's win rate for the delivery robot ranged from 25\% on the secondary road to 57\% on the local road, while that for the humanoid ranged from 26\% to 32\%.
Overall, win rates varied less across road classes than between day and night, with no consistent pattern shared by the two models.

\subsection{Perceived Safety across Persona Attributes}
\label{sec:res_persona}

We next examine how model judgments vary across four persona attributes: gender, age, income, and residential region.
\Cref{fig:persona}a shows opposite patterns in humanoid win rates by gender across the two models.
Against the no-robot condition, female personas chose the humanoid more often than male personas in Gemma (31\% vs.\ 26\%), whereas male personas chose it more often in Qwen (14\% vs.\ 12\%).
Gender differences were smaller for the delivery robot, with female personas choosing it slightly more often in both models.
For income and residential region, \Cref{fig:persona}c and \Cref{fig:persona}d show relatively small differences in win rates across both models.

Age showed a more consistent pattern across both models, as shown in \Cref{fig:persona}b.
From the youngest to the oldest age group, the delivery robot's win rate against the no-robot condition decreased from 54\% to 26\% for Qwen and from 51\% to 28\% for Gemma.
A similar decline appeared for the humanoid only in Gemma, from 39\% to 18\%, whereas Qwen's win rates remained between 9\% and 16\%.
Overall, older personas were less likely to choose the delivery robot over the no-robot condition, while a similar age-related decline for the humanoid appeared only in Gemma.

\subsection{Rationale Topics across Choices and Conditions}
\label{sec:res_rationale}

To examine how models explain their choices, we analyze the rationales of Gemma, which showed a particularly large nighttime increase in preference for the humanoid.
\Cref{fig:rationale} shows the rationale topic distributions across robot choices, time of day, and persona gender.
The rationale topic results for Qwen are detailed in \cref{app:qwen_rationale}.

The two robots were described in different terms.
The delivery robot was mainly described as a physical obstacle, with concerns about blocked sidewalks and tripping hazards.
Collision or path obstruction accounted for 74\% of the rationale topics in this comparison during the day.
In contrast, the humanoid was described as either threatening or protective.
When the no-robot condition was chosen over the humanoid, threat or unfamiliarity represented 73\% of the rationale topics by day and 94\% at night.
When the humanoid was chosen, surveillance, accompaniment, or protection represented more than 90\% of the rationale topics both by day and at night.
Therefore, the same humanoid was described as threatening when the no-robot condition was chosen but as protective when the humanoid was chosen.

As shown in \Cref{fig:rationale}a, topics describing the robot as a help (surveillance and robot function) were more common at night when the delivery robot was chosen, with surveillance increasing from 7\% to 53\%.
\Cref{fig:rationale}b shows differences in rationale topics by persona gender.
When the delivery robot was chosen, help topics were more common for female personas, whereas topics describing the robot as a risk (collision and threat) were more common for male personas.
For the humanoid, surveillance dominated for both genders.
Thus, gender differences in rationale topics were more pronounced for the delivery robot than for the humanoid.

\subsection{Alignment between Human and Model Judgments}
\label{sec:res_alignment}

\begin{figure}[t]
  \centering
  \includegraphics[width=\columnwidth]{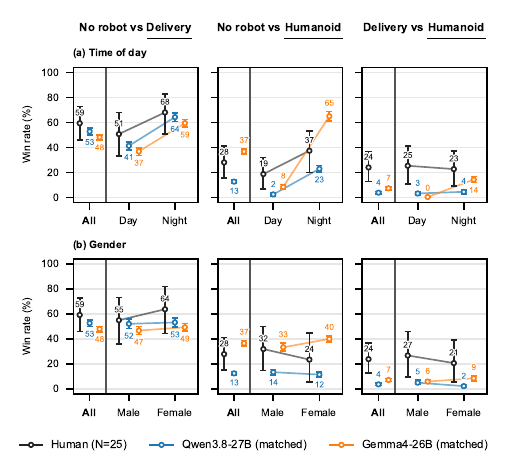}
  \caption{Win rate of humans and matched personas. Rows show (a) time of day and (b) gender, and columns show robot pairs. Model results use personas matched to the human respondents by gender and age group.}
  \label{fig:human_matched}
\end{figure}

We then examine whether the model judgments agree with those of human respondents.
For this comparison, we used personas matched to the respondents by gender and age group.
As shown in \Cref{fig:human_matched}, humans and the models made relatively similar choices when each robot was compared with the no-robot condition.
Humans chose the delivery robot in 59\% of these trials (53\% for Qwen and 48\% for Gemma) and the humanoid in 28\% (13\% for Qwen and 37\% for Gemma).
The models differed from humans when the two robots were compared.
Humans chose the humanoid in 24\% of these trials, but Qwen and Gemma chose it in only 4\% and 7\%.
Thus, humans and the models differed most when choosing between the two robots.

As shown in \Cref{fig:human_matched}a, all three groups chose robots over the no-robot condition more often at night than during the day. When the delivery robot was compared with the no-robot condition, humans chose the delivery robot in 51\% of daytime trials and 68\% of nighttime trials, and both models showed a similar rise.
When the humanoid was compared with the no-robot condition, humans chose the humanoid in 19\% of daytime trials and 37\% of nighttime trials, and Qwen showed a similar rise.
Gemma showed a much larger rise, from 8\% by day to 65\% at night.
The survey responses help explain why humans chose the humanoid at night less often than Gemma did.
Some respondents felt safer with a robot on a dark street, but others expressed concerns about an unfamiliar humanoid or a possible malfunction, as detailed in Appendix~\ref{app:free}.

\Cref{fig:human_matched}b compares female and male respondents.
Female respondents chose the humanoid slightly less often than male respondents (24\% vs.\ 32\% against the no-robot condition and 21\% vs.\ 27\% against the delivery robot).
Qwen showed the same difference, but female personas in Gemma chose the humanoid more often than male personas.
Gender differences were small, as shown by the effect sizes in Appendix~\ref{app:effects}.
Overall, both models followed humans in choosing robots more often at night.
The models differed mainly when choosing between the two robots.
Gemma also showed a much larger nighttime rise for the humanoid than humans did.

\subsection{Response Consistency}
\label{sec:res_consistency}

\begin{table}[!b]
  \centering
  \caption{Response consistency of the two models. Each value shows the proportion of repeated trials where the model chose the same image again. The two images were presented in either the same or swapped positions.}
  \label{tab:consistency}
  \small
  \setlength{\tabcolsep}{3.5pt}
  \begin{tabular}{lcccc}
    \toprule
    & \multicolumn{2}{c}{Qwen3.8-27B} & \multicolumn{2}{c}{Gemma4-26B} \\
    \cmidrule(lr){2-3} \cmidrule(lr){4-5}
    Robot pair & Same & Swapped & Same & Swapped \\
    \midrule
    No robot vs. Delivery & 65\% & 52\% & 71\% & 61\% \\
    No robot vs. Humanoid & 84\% & 77\% & 81\% & 73\% \\
    Delivery vs. Humanoid & 93\% & 93\% & 94\% & 92\% \\
    \midrule
    All pairs & 81\% & 74\% & 82\% & 75\% \\
    \bottomrule
  \end{tabular}
\end{table}

To assess the reliability of the model responses, we compared repeated trials with the original responses.
As shown in \Cref{tab:consistency}, when a trial was repeated with the two images in the same positions as before, Qwen and Gemma chose the same condition in 81\% and 82\% of trials on average.
When the two images were swapped between left and right, consistency decreased to 74\% and 75\%.
Across robot pairs, consistency was highest for the delivery robot against the humanoid (92\% to 94\%) and lowest for the delivery robot against the no-robot condition (52\% to 71\%).
This matches the win rates in \Cref{fig:context}, as choices were more stable when one condition was chosen in most trials (95\%) than when both were chosen at similar rates (45\% and 41\%).
Overall, the model responses were stable mainly when one condition was clearly preferred.

\section{Conclusion}

This study examined how robots affect the perceived safety of streets that pedestrians pass without interacting with them.
We generated street view images that varied the robot, the time of day, and the road class.
We then compared the judgments of two persona-conditioned MLLMs with those of human respondents and analyzed the rationales behind the model choices.
The models generally judged streets without robots as safest and the delivery robot as safer than the humanoid, but chose robots more often at night than by day.
Their rationales showed that the two robots raised different issues, with the delivery robot seen mainly as a physical obstacle and the humanoid as either a threat or a source of protection.
Humans showed the same increase at night but chose the humanoid over the delivery robot far more often than the models did.
Our findings support the view that whether a robot fits a place depends on its context.
Persona-conditioned MLLMs can help explore such effects of context, but their judgments of a specific robot may still differ from those of people.

This study has three main limitations that point to future work.
First, the human sample was small and included only respondents in their 20s, 30s, and 50s, and the images covered three scenes in one city, so the findings should be tested with larger and more diverse samples and with more scenes, cities, and robot types.
Second, the images were static and did not show robots performing their tasks or interacting with people, although seeing a robot deliver a parcel or help someone may change how safe a street feels, so future work should examine robots in action through videos or field studies.
Third, the model results came from only two open-weight models and were partly affected by image position, so future work should reduce such position effects and compare a wider range of models, including closed models.

\printbibliography

@String { iros        = {IEEE/RSJ International Conference on Intelligent Robots and Systems (IROS)} }

@article{salesses2013collaborative,
  title={The collaborative image of the city: mapping the inequality of urban perception},
  author={Salesses, Philip and Schechtner, Katja and Hidalgo, C{\'e}sar A},
  journal={PloS one},
  volume={8},
  number={7},
  pages={e68400},
  year={2013},
  publisher={Public Library of Science San Francisco, USA}
}

@inproceedings{naik2014streetscore,
  title={Streetscore-predicting the perceived safety of one million streetscapes},
  author={Naik, Nikhil and Philipoom, Jade and Raskar, Ramesh and Hidalgo, C{\'e}sar},
  booktitle={Proceedings of the IEEE conference on computer vision and pattern recognition workshops},
  pages={779--785},
  year={2014}
}

@inproceedings{dubey2016deep,
  title={Deep learning the city: Quantifying urban perception at a global scale},
  author={Dubey, Abhimanyu and Naik, Nikhil and Parikh, Devi and Raskar, Ramesh and Hidalgo, C{\'e}sar A},
  booktitle={European conference on computer vision},
  pages={196--212},
  year={2016},
  organization={Springer}
}

@article{li2022measuring,
  title={Measuring visual walkability perception using panoramic street view images, virtual reality, and deep learning},
  author={Li, Yunqin and Yabuki, Nobuyoshi and Fukuda, Tomohiro},
  journal={Sustainable Cities and Society},
  volume={86},
  pages={104140},
  year={2022},
  publisher={Elsevier}
}

@article{cui2023analysing,
  title={Analysing gender differences in the perceived safety from street view imagery},
  author={Cui, Qinyu and Zhang, Yan and Yang, Guang and Huang, Yiting and Chen, Yu},
  journal={International journal of applied earth observation and geoinformation},
  volume={124},
  pages={103537},
  year={2023},
  publisher={Elsevier}
}

@article{xiao2025urban,
  title={How are urban design qualities associated with perceived walkability? An {AI} approach using street view images and deep learning},
  author={Xiao, Yang and Song, Mengkun},
  journal={International Journal of Urban Sciences},
  volume={29},
  number={sup1},
  pages={66--91},
  year={2025},
  publisher={Taylor \& Francis}
}

@article{lefosse2025feeling,
  title={Feeling Nature: Measuring perceptions of biophilia across global biomes using visual {AI}},
  author={Lefosse, Deborah C and Duarte, F{\'a}bio and Sanatani, Rohit Priyadarshi and Kang, Yuhao and van Timmeren, Arjan and Ratti, Carlo},
  journal={npj Urban Sustainability},
  volume={5},
  number={1},
  pages={4},
  year={2025},
  publisher={Nature Publishing Group UK London}
}

@article{ye2025generated,
  title={Generated nighttime street view image to inform perceived safety divergence between day and night in high density cities: A case study in {Hong Kong}},
  author={Ye, Xiaotong and Wang, Yuankai and Dai, Jiajing and Qiu, Waishan},
  journal={Journal of Urban Management},
  volume={14},
  number={2},
  pages={379--401},
  year={2025},
  publisher={Elsevier}
}

@article{dai2026urban,
  title={The urban greening paradox: Measuring day-night perceived safety divergences from paired street view images using generative {AI}},
  author={Dai, Jiajing and Wang, Yuankai and Qiu, Waishan and Ma, Jun and Chen, Da},
  journal={Urban Forestry \& Urban Greening},
  pages={129350},
  year={2026},
  publisher={Elsevier}
}

@inproceedings{ro2025well,
  title={How well do vision-language models understand cities? a comparative study on spatial reasoning from street-view images},
  author={Ro, Juneyoung and Kim, Namwoo and Yoon, Yoonjin},
  booktitle={2025 IEEE/CVF International Conference on Computer Vision Workshops (ICCVW)},
  pages={6535--6544},
  year={2025},
  organization={IEEE}
}

@inproceedings{pelikan2025making,
  title={Making sense of public space for robot design},
  author={Pelikan, Hannah RM and Mutlu, Bilge and Reeves, Stuart},
  booktitle={2025 20th ACM/IEEE International Conference on Human-Robot Interaction (HRI)},
  pages={152--162},
  year={2025},
  organization={IEEE}
}

@inproceedings{dobrosovestnova2025beyond,
  title={Beyond the user: Mapping subject positions for robots in public spaces},
  author={Dobrosovestnova, Anna and Babel, Franziska and Pelikan, Hannah},
  booktitle={2025 20th ACM/IEEE International Conference on Human-Robot Interaction (HRI)},
  pages={163--173},
  year={2025},
  organization={IEEE}
}

@inproceedings{abrams2021theoretical,
  title={A theoretical and empirical reflection on technology acceptance models for autonomous delivery robots},
  author={Abrams, Anna MH and Dautzenberg, Pia SC and Jakobowsky, Carla and Ladwig, Stefan and Rosenthal-Von Der P{\"u}tten, Astrid M},
  booktitle={Proceedings of the 2021 ACM/IEEE International Conference on Human-Robot Interaction},
  pages={272--280},
  year={2021}
}

@article{kruse2013human,
  title={Human-aware robot navigation: A survey},
  author={Kruse, Thibault and Pandey, Amit Kumar and Alami, Rachid and Kirsch, Alexandra},
  journal={Robotics and Autonomous Systems},
  volume={61},
  number={12},
  pages={1726--1743},
  year={2013},
  publisher={Elsevier}
}

@article{nahum2026advancing,
  title={Advancing a taxonomy of proxemics for socially aware robot navigation},
  author={Nahum, Ehud and Edan, Yael and Oron-Gilad, Tal},
  journal={Frontiers in Robotics and AI},
  volume={13},
  pages={1800762},
  year={2026},
  publisher={Frontiers Media SA}
}

@article{gehrke2023observed,
  title={Observed sidewalk autonomous delivery robot interactions with pedestrians and bicyclists},
  author={Gehrke, Steven R and Phair, Christopher D and Russo, Brendan J and Smaglik, Edward J},
  journal={Transportation research interdisciplinary perspectives},
  volume={18},
  pages={100789},
  year={2023},
  publisher={Elsevier}
}

@inproceedings{aher2023using,
  title={Using large language models to simulate multiple humans and replicate human subject studies},
  author={Aher, Gati V and Arriaga, Rosa I and Kalai, Adam Tauman},
  booktitle={International conference on machine learning},
  pages={337--371},
  year={2023},
  organization={PMLR}
}

@article{argyle2023out,
  title={Out of one, many: Using language models to simulate human samples},
  author={Argyle, Lisa P and Busby, Ethan C and Fulda, Nancy and Gubler, Joshua R and Rytting, Christopher and Wingate, David},
  journal={Political Analysis},
  volume={31},
  number={3},
  pages={337--351},
  year={2023},
  publisher={Cambridge University Press}
}

@inproceedings{zhang2023large,
  title={Large language models as zero-shot human models for human-robot interaction},
  author={Zhang, Bowen and Soh, Harold},
  booktitle={2023 IEEE/RSJ International Conference on Intelligent Robots and Systems (IROS)},
  pages={7961--7968},
  year={2023},
  organization={IEEE}
}

@inproceedings{hu2024quantifying,
  title={Quantifying the persona effect in {LLM} simulations},
  author={Hu, Tiancheng and Collier, Nigel},
  booktitle={Proceedings of the 62nd Annual Meeting of the Association for Computational Linguistics (Volume 1: Long Papers)},
  pages={10289--10307},
  year={2024}
}

@inproceedings{meister2025benchmarking,
  title={Benchmarking distributional alignment of large language models},
  author={Meister, Nicole and Guestrin, Carlos and Hashimoto, Tatsunori B},
  booktitle={Proceedings of the 2025 Conference of the Nations of the Americas Chapter of the Association for Computational Linguistics: Human Language Technologies (Volume 1: Long Papers)},
  pages={24--49},
  year={2025}
}

@article{verma2026generative,
  title={Generative agents in the streets: Exploring the use of Large Language Models ({LLMs}) in collecting urban perceptions},
  author={Verma, Deepank and Mumm, Olaf and Carlow, Vanessa Miriam},
  journal={Computers in Human Behavior: Artificial Humans},
  pages={100277},
  year={2026},
  publisher={Elsevier}
}

@article{gu2026survey,
  title={A survey on {LLM}-as-a-judge},
  author={Gu, Jiawei and Jiang, Xuhui and Shi, Zhichao and Tan, Hexiang and Zhai, Xuehao and Xu, Chengjin and Li, Wei and Shen, Yinghan and Ma, Shengjie and Liu, Honghao and others},
  journal={The Innovation},
  volume={7},
  number={6},
  year={2026},
  publisher={Elsevier}
}

@inproceedings{neuhold2017mapillary,
  title={The {Mapillary Vistas} dataset for semantic understanding of street scenes},
  author={Neuhold, Gerhard and Ollmann, Tobias and Rota Bulo, Samuel and Kontschieder, Peter},
  booktitle={Proceedings of the IEEE international conference on computer vision},
  pages={4990--4999},
  year={2017}
}

@article{gilardi2023chatgpt,
  title={{ChatGPT} outperforms crowd workers for text-annotation tasks},
  author={Gilardi, Fabrizio and Alizadeh, Meysam and Kubli, Ma{\"e}l},
  journal={Proceedings of the National Academy of Sciences},
  volume={120},
  number={30},
  pages={e2305016120},
  year={2023},
  publisher={National Academy of Sciences}
}

@inproceedings{pelikan2024encountering,
  title={Encountering autonomous robots on public streets},
  author={Pelikan, Hannah RM and Reeves, Stuart and Cantarutti, Marina N},
  booktitle={Proceedings of the 2024 ACM/IEEE International Conference on Human-Robot Interaction},
  pages={561--571},
  year={2024}
}

@article{biederman1982scene,
  title={Scene perception: Detecting and judging objects undergoing relational violations},
  author={Biederman, Irving and Mezzanotte, Robert J and Rabinowitz, Jan C},
  journal={Cognitive psychology},
  volume={14},
  number={2},
  pages={143--177},
  year={1982},
  publisher={Elsevier}
}

@article{loftus1978cognitive,
  title={Cognitive determinants of fixation location during picture viewing},
  author={Loftus, Geoffrey R and Mackworth, Norman H},
  journal={Journal of Experimental Psychology: Human perception and performance},
  volume={4},
  number={4},
  pages={565--572},
  year={1978},
  publisher={American Psychological Association}
}

@article{reber2004processing,
  title={Processing fluency and aesthetic pleasure: Is beauty in the perceiver's processing experience?},
  author={Reber, Rolf and Schwarz, Norbert and Winkielman, Piotr},
  journal={Personality and social psychology review},
  volume={8},
  number={4},
  pages={364--382},
  year={2004},
  publisher={Sage Publications Sage CA: Los Angeles, CA}
}

@inproceedings{kwon2023efficient,
  title={Efficient memory management for large language model serving with {PagedAttention}},
  author={Kwon, Woosuk and Li, Zhuohan and Zhuang, Siyuan and Sheng, Ying and Zheng, Lianmin and Yu, Cody Hao and Gonzalez, Joseph and Zhang, Hao and Stoica, Ion},
  booktitle={Proceedings of the 29th symposium on operating systems principles},
  pages={611--626},
  year={2023}
}

@book{cohen1988statistical,
  title={Statistical power analysis for the behavioral sciences},
  author={Cohen, Jacob},
  edition={2},
  year={1988},
  publisher={Lawrence Erlbaum Associates},
  address={Hillsdale, NJ}
}

@article{gemmateam2026gemma4,
  title={{Gemma 4} technical report},
  author={{Gemma Team}},
  journal={arXiv preprint arXiv:2607.02770},
  year={2026}
}

@misc{molit_road_classification,
  title={Rules on the Decision, Structure, and Installation Standards of Urban and county Planning Facilities, Article 9, Item 3},
  author={{Ministry of Land, Infrastructure and Transport}},
  howpublished={Korean Law Information Center, \url{https://www.law.go.kr}},
  note={In Korean; English title translated by the authors. Version in force as of Oct. 14, 2026}
}

@misc{openai2026gpt56sol,
  title={{GPT-5.6 Sol}},
  author={{OpenAI}},
  howpublished={\url{https://developers.openai.com/api/docs/models/gpt-5.6-sol}},
  note={Accessed Oct. 14, 2026}
}

@misc{qwen2026qwen38_27b,
  title={{Qwen3.8-27B}},
  author={{Qwen Team}},
  howpublished={\url{https://huggingface.co/Qwen/Qwen3.8-27B}},
  note={Hugging Face model card. Accessed Oct. 16, 2026}
}

@misc{anthropic2026opus55,
  title={System card: {Claude Opus 5.5}},
  author={{Anthropic}},
  howpublished={\url{https://www-cdn.anthropic.com/fc1b44717c85dc068bc6ba5024219938094694bd/Claude\%20Opus\%205.5\%20System\%20Card.pdf}},
  note={Accessed Oct. 16, 2026}
}

@misc{nvidia2026nemotronkorea,
  title={{Nemotron-Personas-Korea}: Synthetic personas aligned to real-world distributions for {Korea}},
  author={Kim, Hyunwoo and Ryu, Jihyeon and Lee, Jinho and Ryu, Hyungon and Praveen, Kiran and Prayaga, Shyamala and Thadaka, Kirit and Jennings, Will and Sadeghi, Bardiya and Sharabiani, Ashton and Choi, Yejin and Meyer, Yev},
  howpublished={\url{https://huggingface.co/datasets/nvidia/Nemotron-Personas-Korea}},
  year={2026}
}

@misc{redhat_qwen38_int4,
  author       = {{Red Hat AI}},
  title        = {{Qwen3.8-27B-INT4}},
  howpublished = {\url{https://huggingface.co/RedHatAI/Qwen3.8-27B-INT4}},
  note         = {Hugging Face model card}
}

@misc{cyankiwi_gemma4_awq,
  author       = {{cyankiwi}},
  title        = {{gemma-4-26B-A4B-it-AWQ-4bit}},
  howpublished = {\url{https://huggingface.co/cyankiwi/gemma-4-26B-A4B-it-AWQ-4bit}},
  note         = {Hugging Face model card}
}

\clearpage
\appendices
\crefalias{section}{appendix}
\raggedbottom

\section{Generated Street View Images}
\label{app:stimuli}

\begin{figure}[H]
  \centering
  \includegraphics[width=\columnwidth]{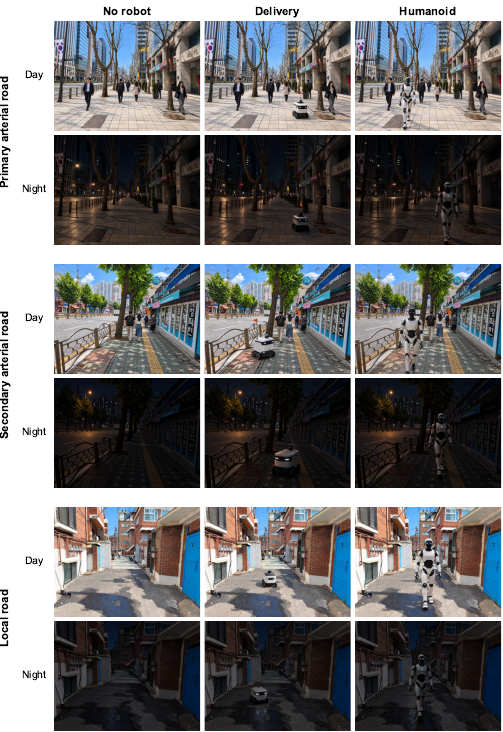}
  \caption{The 18 stimulus images. Rows show the three road classes by day and at night, and columns show robot conditions. Each trial compared two images in the same row.}
  \label{fig:stimuli}
\end{figure}

\Cref{fig:stimuli} shows all 18 images. Each row shows one road class at one time of day, and each column shows one robot condition. All images were resized to $1024 \times 768$ pixels. Few pedestrians were reinserted in the night images, reflecting typical activity at midnight.

\newpage

\section{Persona Cards}
\label{app:personas}
Each persona card lists five attributes, followed by a one-line introduction and a short description of household income. The cards were given to the models in Korean; the two examples below are translated by the authors. \Cref{tab:personas} shows the composition of all personas and of the personas matched to the human respondents.

\medskip
\noindent\fbox{\parbox{\dimexpr\columnwidth-2\fboxsep-2\fboxrule}{\small
Gender: Female\\
Age: 25 (20s or younger)\\
Occupation: Unemployed\\
Household income level: Low\\
Residence: Michuhol-gu, Incheon (capital region)\\[2pt]
Ms.\ Chu Yun-ji is a 25-year-old job seeker in Incheon with strong artistic sensitivity and empathy but weak practical self-management.\\[2pt]
Her household income is low, and because she is preparing for employment, housing costs outside her current home, transportation, and food take up a large share of living expenses. She saves in advance for large expenses such as exam fees and interview clothes and manages within a fixed budget.
}}

\begin{table}[H]
  \centering
  \caption{Composition of the personas. All shows the 500 personas used for the model results, and Matched shows the 137 unique personas matched to the human respondents.}
  \label{tab:personas}
  \begin{tabular}{llrr}
    \toprule
    Attribute & Level & All & Matched \\
    \midrule
    Gender & Female & 250 & 75 \\
     & Male & 250 & 62 \\
    \midrule
    Age group & 20s or younger & 84 & 74 \\
     & 30s & 84 & 39 \\
     & 40s & 83 & 0 \\
     & 50s & 83 & 24 \\
     & 60s & 83 & 0 \\
     & 70s or older & 83 & 0 \\
    \midrule
    Household income & Low & 167 & 46 \\
     & Middle & 166 & 43 \\
     & High & 167 & 48 \\
    \midrule
    Region & Capital region & 250 & 77 \\
     & Other regions & 250 & 60 \\
    \bottomrule
  \end{tabular}
\end{table}

\newpage
\section{Rationale Coding}
\label{app:coding}

Each rationale received up to three topics, regardless of direction: for example, both a high and a low risk of collision were coded as topic (1). Each of the five topics is defined below and illustrated with one rationale that received only that topic, translated from Korean.

{\small
\smallskip
\noindent\textbf{(1) Collision or path obstruction.} The robot may collide with pedestrians, cause them to trip, or block the walking path. \emph{Example (Gemma, chose no robot, day):} ``A small delivery robot on the walkway could disrupt the flow of pedestrians or bump into them, so on a crowded sidewalk the right image without the robot feels safer.''

\smallskip
\noindent\textbf{(2) Threat or unfamiliarity.} The size, human-like form, or approach of the robot feels threatening, its movement seems unpredictable, or it looks unfamiliar. \emph{Example (Gemma, chose no robot, day):} ``A humanoid robot in human form may feel threatening or unsettling because of its unfamiliar movements, so the right image without the robot feels much calmer and safer.''

\smallskip
\noindent\textbf{(3) Openness, visibility, or lighting.} The robot blocks or does not block the view, the path is open, the robot is easy to see and avoid, or its light illuminates the path. \emph{Example (Gemma, chose the delivery robot, night):} ``When passing through a dark alley, the light from the small delivery robot makes me less anxious and makes it safer to check the condition of the path.''

\smallskip
\noindent\textbf{(4) Surveillance, accompaniment, or protection.} The presence of the robot is reassuring because it watches or patrols the street, signals a managed area, gives a sense that someone is nearby, or may help in an emergency. \emph{Example (Gemma, chose the delivery robot, day):} ``Walking alone in a narrow, quiet alley can feel a little uneasy, but the delivery robot gives a sense of movement nearby, so it feels psychologically safer.''

\smallskip
\noindent\textbf{(5) Robot function or trust in technology.} The robot is harmless because it is a simple machine that only carries items, or it is trustworthy because its technology can be controlled. \emph{Example (Qwen, chose the delivery robot, night):} ``As an electronics engineer, I trust the technical safety of small robots, so a space with a small delivery robot feels much safer and calmer than one with a large humanoid robot.''
\par
}

\smallskip
\Cref{tab:kappa} shows the agreement between the human annotator and Claude on the 150 validation rationales described in \cref{sec:method_rationale}. The annotator did not assign topic (5) to any of these rationales, so its $\kappa$ is not defined.

\begin{table}[H]
  \centering
  \caption{Agreement between the human annotator and Claude for each topic. Counts give the number of rationales that each coder assigned to the topic and sum to more than 150 because a rationale could receive more than one topic.}
  \label{tab:kappa}
  \small
  \setlength{\tabcolsep}{4pt}
  \begin{tabular}{lrrr}
    \toprule
    Topic & Annotator & Claude & $\kappa$ \\
    \midrule
    (1) Collision or path obstruction & 70 & 71 & 0.83 \\
    (2) Threat or unfamiliarity & 100 & 104 & 0.82 \\
    (3) Openness, visibility, or lighting & 20 & 23 & 0.81 \\
    (4) Surveillance or protection & 33 & 29 & 0.92 \\
    (5) Robot function or trust in technology & 0 & 5 & -- \\
    \midrule
    All topics & 223 & 232 & 0.87 \\
    \bottomrule
  \end{tabular}
\end{table}

\newpage
\section{Rationale Topics of Qwen}
\label{app:qwen_rationale}

\begin{figure}[H]
  \centering
  \includegraphics[width=\columnwidth]{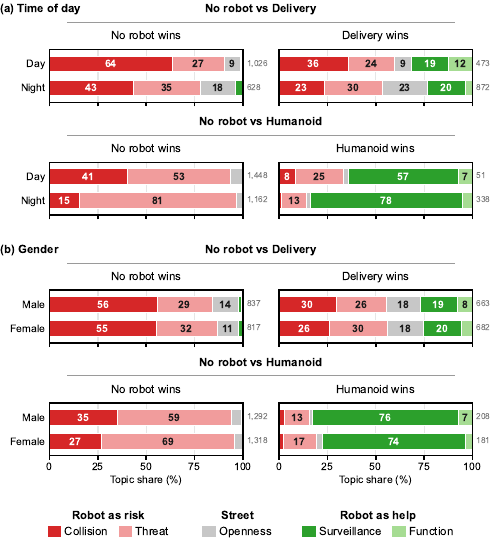}
  \caption{Share of rationale topics stated by Qwen3.8-27B for each choice, by (a) time of day and (b) persona gender. Shares within each bar sum to 100\%, and numbers to the right give the number of responses.}
  \label{fig:rationale_qwen}
\end{figure}

\Cref{fig:rationale_qwen} shows that the rationales of Qwen described the two robots in a way similar to those of Gemma in \cref{sec:res_rationale}.
When Qwen chose the no-robot condition over the delivery robot, 64\% of the rationale topics by day concerned collision or path obstruction, showing that the delivery robot was seen mainly as a physical obstacle.
The humanoid was seen in two opposite ways.
Rationales for avoiding the humanoid mostly cited threat or unfamiliarity (53\% by day and 81\% at night), whereas rationales for choosing the humanoid mostly cited surveillance (57\% by day and 78\% at night).

\Cref{fig:rationale_qwen}a shows that time of day had little effect on the rationales for choosing the delivery robot, as help topics stayed at 19\% by day and 20\% at night.
At night, openness, visibility, or lighting appeared more often instead.
\Cref{fig:rationale_qwen}b shows that persona gender also made little difference.
Female and male personas gave similar help and risk topics for the delivery robot, and both mostly cited surveillance for the humanoid.

\newpage
\section{Free Responses of Human Respondents}
\label{app:free}

15 of the 25 respondents described what mattered most in their safety judgments. Their responses, translated from Korean by the authors, are listed below in two groups. Gender and age group are given in parentheses.

\medskip
\noindent\textbf{Robots as a physical risk or an unfamiliar presence}
\begin{itemize}\small
\item ``The road was very narrow, so having a robot there felt somewhat burdensome. It felt like it was blocking the way, so it did not feel safe.'' (Male, 20s)
\item ``Because it is a robot, it could malfunction, so I worry about bumping into it up close. For the humanoid in particular, I saw on YouTube that it moves dangerously, like a seizure, when it falls, so that worries me.'' (Male, 30s)
\item ``I judged that the delivery robot was unlikely to injure people, but the humanoid could cause harm through malfunction, so meeting it in the middle of the night would not feel safe.'' (Male, 20s)
\item ``I mainly considered how predictable the robot's behavior was and how unclear its purpose was. I worried about collisions because it was hard to predict where the humanoid and the delivery robot would move. I felt especially uneasy about the humanoid because it was hard to know its purpose and role.'' (Female, 30s)
\item ``Humanoid robots do not feel familiar yet.'' (Female, 50s)
\item ``It would be very scary to meet a humanoid on a street at night.'' (Female, 20s)
\item ``I chose the one that did not feel unfamiliar, the one that did not feel scary.'' (Female, 20s)
\item ``I judged that it was safe without a robot. If a robot had to be there, I considered whether it was a machine-like robot that did not resemble a human.'' (Male, 30s)
\end{itemize}

\noindent\textbf{Robots as protection, depending on the context}
\begin{itemize}\small
\item ``During the day, it seems more natural and safer without a robot, but at night, I think a robot could actually make the street safer.'' (Male, 20s)
\item ``I mainly compared safety based on my first impression of each photo. At night, when the street felt dangerous, a street with a robot felt safe, as if there were something to rely on. During the day, when the street felt safe, the robot felt like a threat instead. In short, my impression of the robot depended on the background.'' (Male, 20s)
\item ``The humanoid was unfamiliar, so it did not feel safe. In the dark, a street with a delivery robot giving off light felt safer than an empty street. In daylight, I did not feel safe because I worried that people walking on the street might bump into the robot.'' (Female, 20s)
\item ``I saw the humanoid as a robot that can actually act, so I mainly considered whether it could help me even a little in a dangerous situation. For the delivery robot, I focused less on security and my own safety and more on whether it would be an obstacle while walking.'' (Male, 30s)
\item ``When I felt that someone was watching over me, a visible robot felt safer.'' (Female, 50s)
\item ``The robot moved freely, and above all, it felt reassuring because it seemed like it would keep people safe.'' (Male, 20s)
\item ``I looked at whether the form and function of the robot fit its surroundings.'' (Male, 20s)
\end{itemize}

\newpage
\section{Effect Sizes of Time of Day and Gender}
\label{app:effects}

\begin{figure}[!htbp]
  \centering
  \includegraphics[width=\columnwidth]{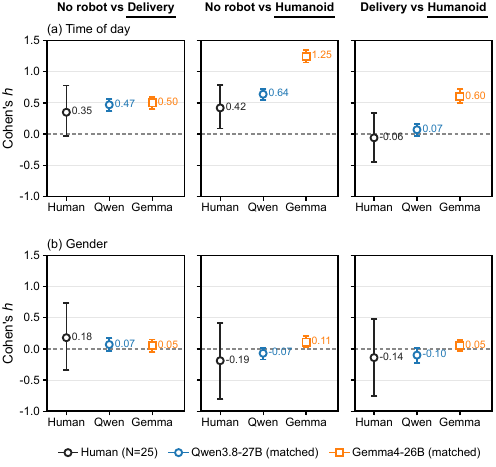}
  \caption{Effect sizes for humans and matched personas. Rows show (a) time of day and (b) gender, and columns show robot pairs. Each value is Cohen's $h$ for the change in the win rate of the underlined target condition, from day to night or from male to female. Error bars show 95\% confidence intervals, and the dashed line marks no effect.}
  \label{fig:effects}
\end{figure}

\Cref{fig:effects} expresses the changes in win rate shown in \Cref{fig:human_matched} as Cohen's $h$.
Positive values indicate a higher win rate at night or for female respondents and personas.
The confidence intervals of humans are wide because the human sample included only 25 respondents.

For time of day, all three groups showed similar effects for the delivery robot against the no-robot condition ($h = 0.35$ for humans, $0.47$ for Qwen, and $0.50$ for Gemma).
For the humanoid against the no-robot condition, the effect of Qwen ($h = 0.64$) stayed within the confidence interval of humans ($h = 0.42$).
The effect of Gemma was large ($h = 1.25$) and fell outside the interval.
Gemma also showed a medium effect for the humanoid against the delivery robot ($h = 0.60$), whereas humans and Qwen showed almost none.

For gender, all effects were below the threshold for a small effect ($|h| < 0.2$).
Humans and Qwen showed negative effects for the humanoid, while Gemma showed positive effects.

\end{document}